\documentclass[twocolumn]{aastex631}

\received{2026 February}
\accepted{2026 X X }

\usepackage{longtable}
\usepackage{graphics}
\usepackage{epsf}
\usepackage{graphicx}	
\usepackage{amsmath}	
\usepackage{amssymb}	
\usepackage{hyperref}
\usepackage{xcolor}

\shorttitle{Third Outburst of IC 3599}
\shortauthors{D. Grupe et al.}

\begin{document}

\title{In the Eye of the Storm: The Third Giant X-ray Outburst of the Extreme Changing-look AGN IC 3599}

 \author[0000-0002-9961-3661]{D. Grupe}
\correspondingauthor{D. Grupe}
\email{dgrupe007@gmail.com}
 \affiliation{
Department of Physics, Geology, and Engineering Technology,
Northern Kentucky University, 
Nunn Drive, 
Highland Heights, KY 41099, USA }

\author[0000-0003-4183-4215]{S. Komossa}
\correspondingauthor{S. Komossa}
\email{skomossa@mpifr.de}
\affiliation{Max-Planck-Institut f{\"u}r Radioastronomie, 
Auf dem H{\"u}gel 69, 53121 Bonn, Germany}

\author[0000-0002-2636-6508]{W. Zheng}
\affiliation{Department of Astronomy, 
University of California, 
 Berkeley, CA 94720-3411, USA}

\author[0000-0003-3460-0103]{A. V. Filippenko}
\affiliation{Department of Astronomy, 
University of California, 
 Berkeley, CA 94720-3411, USA}

\author[0000-0001-5955-2502]{T. G. Brink}
\affiliation{Department of Astronomy, %
 University of California, 
 Berkeley, CA 94720-3411, USA}

\author{N. Schartel}
\affiliation{EMFTEL Department, Universidad Complutense de Madrid, Plaza de Ciencias,
1. Ciudad Universitaria, E-28040 Madrid, Spain.}

\author{J. Wang}
\affiliation{Key Laboratory of Space Astronomy and Technology, National Astronomical Observatories, Chinese Academy of Sciences, Beijing 100101,
People's Republic of China}

\author{V. Oknyansky}
\affiliation{Sternberg Astronomical Institute, M.V. Lomonosov Moscow State University,  119234, Moscow, Universitetsky pr-t, 13, Russia}
\affiliation{Department of Physics, Faculty of Natural Sciences, University of Haifa, Haifa 3498838, Israel}

\author{A. V. Dodin}
\affiliation{Sternberg Astronomical Institute, M.V. Lomonosov Moscow State University,  119234, Moscow, Universitetsky pr-t, 13, Russia}

\author{S. Wolsing}
\affiliation{Institut für Astrophysik und Geophysik, University of G\"ottingen, 
Friedrich-Hund-Platz 1,
37077 G\"ottingen, Germany}

\author{E. Elien}
\affiliation{Department of Physics \& Astronomy
University of California, Los Angeles
430 Portola Plaza, Knudsen Hall 4-00A
Los Angeles, CA 90095-1547, USA}

\author{C. Tapper}
\affiliation{Department of Physics \& Astronomy
University of California, Los Angeles
430 Portola Plaza, Knudsen Hall 4-00A
Los Angeles, CA 90095-1547, USA}

\author{P. Lynam}
\affiliation{UCO/Lick Observatory, University of California,
Santa Cruz, CA 95064, USA}

\newcommand{\swift}{{\it Swift}}
\newcommand{\suzaku}{{\it Suzaku}}
\newcommand{\xmm}{{\it XMM-Newton}}
\newcommand{\chandra}{{\it Chandra}}
\newcommand{\ax}{$\alpha_{\rm X}$}
\newcommand{\rb}[1]{\raisebox{1.5ex}[-1.5ex]{#1}}
\newcommand{\msun}{$M_{\odot}$}
\newcommand{\dM}{\dot M}
\newcommand{\dMM}{$\dot{M}/M$}
\newcommand{\dMedd}{\dot M_{\rm Edd}}
\newcommand{\plm}{$\pm$}
\newcommand{\nh}{$N_{\rm H}$}
\newcommand{\auv}{$\alpha_{\rm UV}$}
\newcommand{\aox}{$\alpha_{\rm ox}$}
\newcommand{\wpvs}{{WPVS~007}}
\newcommand{\lledd}{$L/L_{\rm Edd}$}
\newcommand{\kms}{km~s$^{-1}$ }

\begin{abstract}
We report the discovery and multiwavelength follow-up observations of a giant (factor $>$100) X-ray outburst of the exceptional changing-look active galactic nucleus (AGN) IC 3599. This is the third such outburst after two previous ones serendipitously discovered in 1990 and 2010. 
Based on our dedicated long-term monitoring of IC 3599 with {\it Swift}, the third outburst was detected while it was happening, and we triggered multiple follow-up observations
within days to weeks for the first time. 
The {\it Swift} outburst spectra are supersoft and almost no photons are detected beyond 2.5 keV. 
The {\it XMM-Newton} short-term light curve shows a remarkable apparent oscillatory pattern that is 
reminiscent of quasiperiodic oscillations (QPOs).
The optical high-state spectra reveal a multitude of bright coronal emission lines that have dramatically brightened and were absent or much fainter in low-state spectra.
The new results eliminate outburst scenarios that require a constant time interval of repetitions (like certain variants of repeat tidal stripping, or of an orbiting supermassive black hole impacting the inner accretion disk), but remain in excellent agreement with an accretion disk radiation-pressure instability when assuming that local conditions in the disk of this long-lived AGN affect the onset time of each new instability.
The combination of recurrent, giant, supersoft outbursts on decadal timescales, the exceptional emission-line response, and the rapid, candidate quasiperiodic, short-term variability on an hours timescale makes IC 3599 unique among AGN, and establishes it as a key system for studying accretion physics under extreme conditions and at the Eddington limit.

\end{abstract}

\keywords{active galactic nuclei -- supermassive black holes  -- X-rays: galaxies -- ultraviolet: galaxies  -- quasars: individual (IC 3599)}

\section{Introduction}
Rapid, high-amplitude X-ray--extreme-ultraviolet (EUV) outbursts in active galactic nuclei (AGN) provide us with important insights in the physics of the accretion disk, and the composition and structure of the illuminated gas in the supermassive black hole's (SMBH) vicinity, including the broad-line region (BLR) and the coronal-line region (CLR).
Some AGN outbursts or deep minimum states are accompanied by changes in the strength of the BLR emission lines, such that in extreme cases Seyfert galaxies can change their optical spectroscopic classification from Seyfert 2 into Seyfert 1 and vice versa. The first such cases were observed in the 1960s--1980s \citep[e.g.,][]{Andrillat1968, Tohline1976, kollatschny1985}. 

Among the growing number of recently identified optical changing-look AGN \citep{Potts2021, Shen2025, Dong2025} \citep[see the reviews by][]{KomossaGrupe2024, ricci2023, Komossa2026A}, IC 3599 \citep[at redshift $z=0.02078$; ][]{rines2016} stands out as an early and extreme example. 
The first X-ray outburst of IC 3599 was discovered
 in 1990 during the {\it ROSAT} all-sky survey \citep[RASS;][]{voges1999}, and its optical spectrum, taken by chance in the high state during the course of identifying {\it ROSAT} EUV sources \citep{Mason1995}, was characterized by bright, broad hydrogen Balmer lines and several strong high-ionization coronal lines \citep{Brandt1995}. The X-ray flux and the emission lines had faded strongly in follow-up observations during the subsequent years \citep{Grupe1995a, KomossaBade1999}, and the optical spectrum in quiescence is that of a Seyfert 1.9 galaxy with only faint, broad H$\alpha$ emission detectable.
IC 3599 was among the AGN with the highest amplitudes of variability in the RASS, and one of the softest X-ray AGN \citep{Brandt1995, grupe1998, grupe2001}.
The high-state emission lines represented a response to the outburst emission of this long-lived Seyfert galaxy, and photoionization modeling of the optical outburst spectrum showed that the properties of the line-emitting region are typical of a CLR with column density $N_{\rm H}=10^{23}$ cm$^{-2}$, density $n=10^9$ cm$^{-3}$, and  a distance from the nucleus of 0.1 pc \citep{KomossaBade1999}. 

 Among different outburst scenarios
an accretion-disk instability was favored \citep{KomossaBade1999}.
A second outburst of IC 3599 of similar amplitude was then discovered by the {\it Neil Gehrels Swift Observatory} ({\it Swift}, hereafter) in 2010, but not immediately noticed such that no high-state follow-up observations were obtained \citep{Campana2015, Grupe2015}. Like during the first outburst, the X-ray spectrum was found to be extremely soft.
Analytic models of an accretion-disk radiation-pressure instability \citep{Lightman1974, Belloni1997} were found to describe the observations well. An observed recurrence time of $\Delta t = 19.5$ yr and an SMBH mass of $10^{6-7}$ $M_{\odot}$ implied a truncation radius between the inner and outer disk of $R_{\rm{trunc}}=$ (5--45)\,$R_{\rm{g}}$. Alternative models, like different variants of rare (repeat) tidal disruption events (TDEs), or binary SMBH scenarios,  \citep{Komossa2014, Campana2015, Grupe2015}, could not yet be fully excluded.

We then set up a monthly monitoring program with {\it Swift} starting in 2013, in search of any new outburst \citep{Grupe2024}, demonstrating that an outburst predicted by the repeat-TDE model in the year 2019/20 was absent. Based on this program, we now report our {\it Swift} discovery of a third factor $>$100 X-ray outburst of IC 3599. 
When \swift\ started observing IC 3599 on 2025 October 31 \citep{Grupe2025} as part of our monthly monitoring campaign, it was found to be more than 50 times brighter than during its low-state level, implying a dramatic flux increase within less than 3 months.
We then started a dense monitoring campaign with \swift\ and triggered multiple follow-up observations.  
Such data were not possible for the previous two outbursts of IC 3599,
since they were only recognized after they had already
faded, so no simultaneous multiwavelength coverage existed in the past.
Here, we report on the {\it Swift} observations, an {\it XMM-Newton} light curve, optical follow-up spectroscopy, and present a discussion of outburst scenarios.

We use a $\Lambda$CDM
cosmology with $\Omega_{\rm M}=0.3$, $\Omega_{\Lambda}=0.7$, and a Hubble
constant H$_0=70$ km s$^{-1}$ Mpc$^{-1}$. At the source's redshift $z=0.02078$ this results in a luminosity
 distance of $D = 91.0$ Mpc 
using the cosmology calculator  by \citet{Wright2006}.
Calendar dates are reported in UTC. 
Uncertainties are 1$\sigma$ unless stated otherwise.

\section{Swift Observations}

\subsection{XRT}
Information on the cadence and duration of observations is provided in Appendix A. 
Data were reduced following standard procedures as described by \cite{Grupe2024}. 
At the start of the monitoring campaign, all observations with the {\it Swift} XRT \citep{Burrows2005} were performed in photon-counting mode \citep{hill2004}. This mode required a pile-up correction for count rates (CR) above 0.6 counts s$^{-1}$. For these observations, an annulus with inner and outer radii of 5 and 40 pixels were used for the source-count extraction, leaving out the inner part of the point-spread function (PSF). Owing to the continued high flux state of IC 3599, we changed the XRT observing mode to windowed timing (WT) starting with the observation on 2025 December 09 in order to avoid pile-up. 

\subsection{UVOT}

{\it Swift}'s ultraviolet-optical telescope \citep[UVOT;][]{Roming2005} was operated in all 6 filters, and the extraction and reduction of the data were performed as described by \cite{Grupe2024}.
The merged UVOT $M2$ and $W2$ images during the current outburst
show that the optical transient's position is consistent with the galaxy's nucleus within the UVOT resolution. We measure UVOT coordinates (J2000) of 
R.A. = $12^{\rm hr}37^{\rm m}41.21^{\rm s}$ in both filters, and 
Dec. = $26^\circ 42' 27.45''$ in $M2$ and Dec. = $26^\circ 42' 27.49''$ in $W2$.
This 
compares to the optical pre-outburst coordinates of the galaxy's nucleus of R.A. = $12^{\rm hr}37^{\rm m}41.22^{\rm s}$ and Dec. = $26^\circ 42' 27.42''$ measured from the 99 ks exposure $W2$ image from \cite{Grupe2024}, 
and corresponds to a projected angular separation of 0.1$''$ and 0.2$''$, respectively.

\begin{figure}
\includegraphics[trim=10 150 15 200,clip,width=9cm]{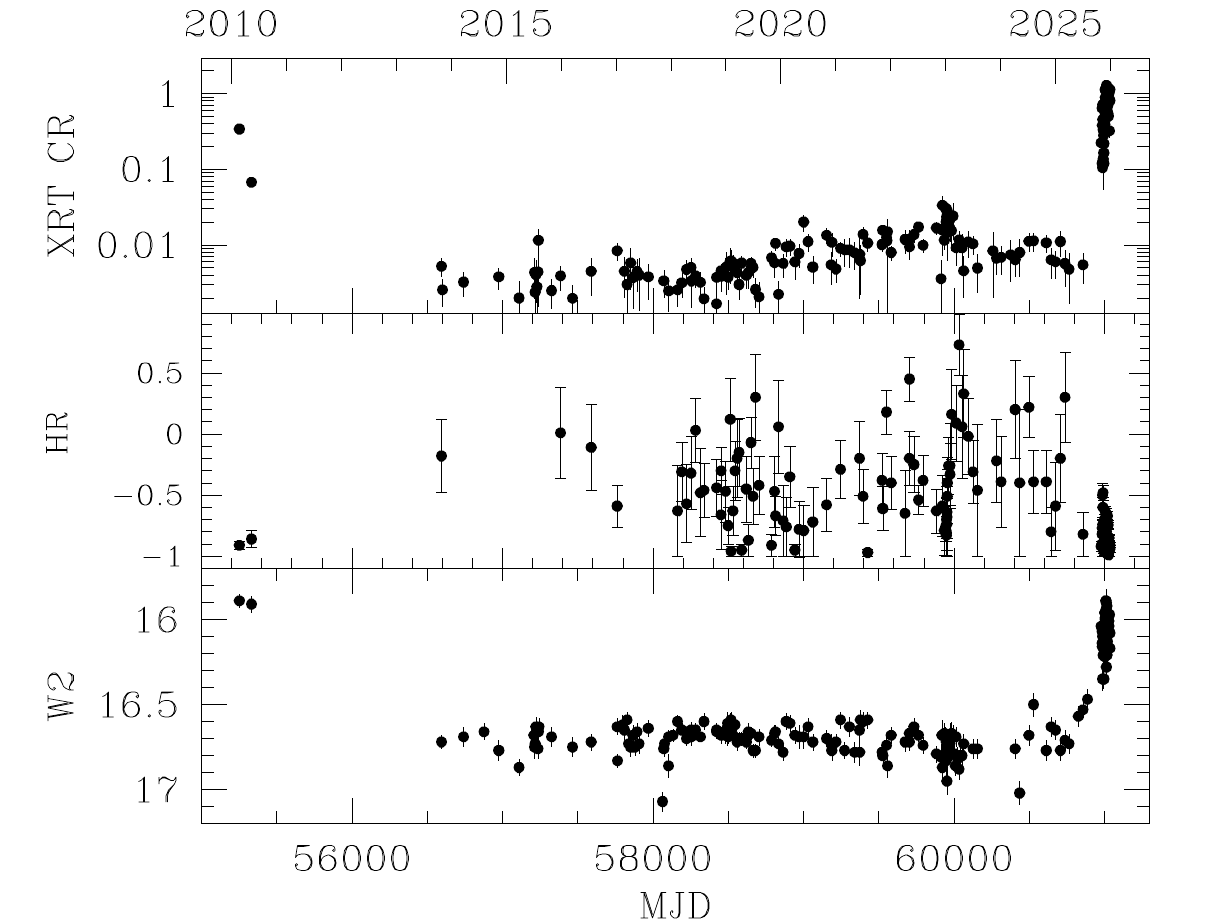}
    \caption{Long-term {\it Swift} XRT count rate, hardness ratio, and UVW2 light curves from 2010 to 2025, including the two high-amplitude outbursts of IC 3599. The third one from 1990 is not shown.   
    }
\label{swift_lc}
\end{figure}

\begin{figure}
\includegraphics[trim=30 150 0 85,clip,width=9.3cm]{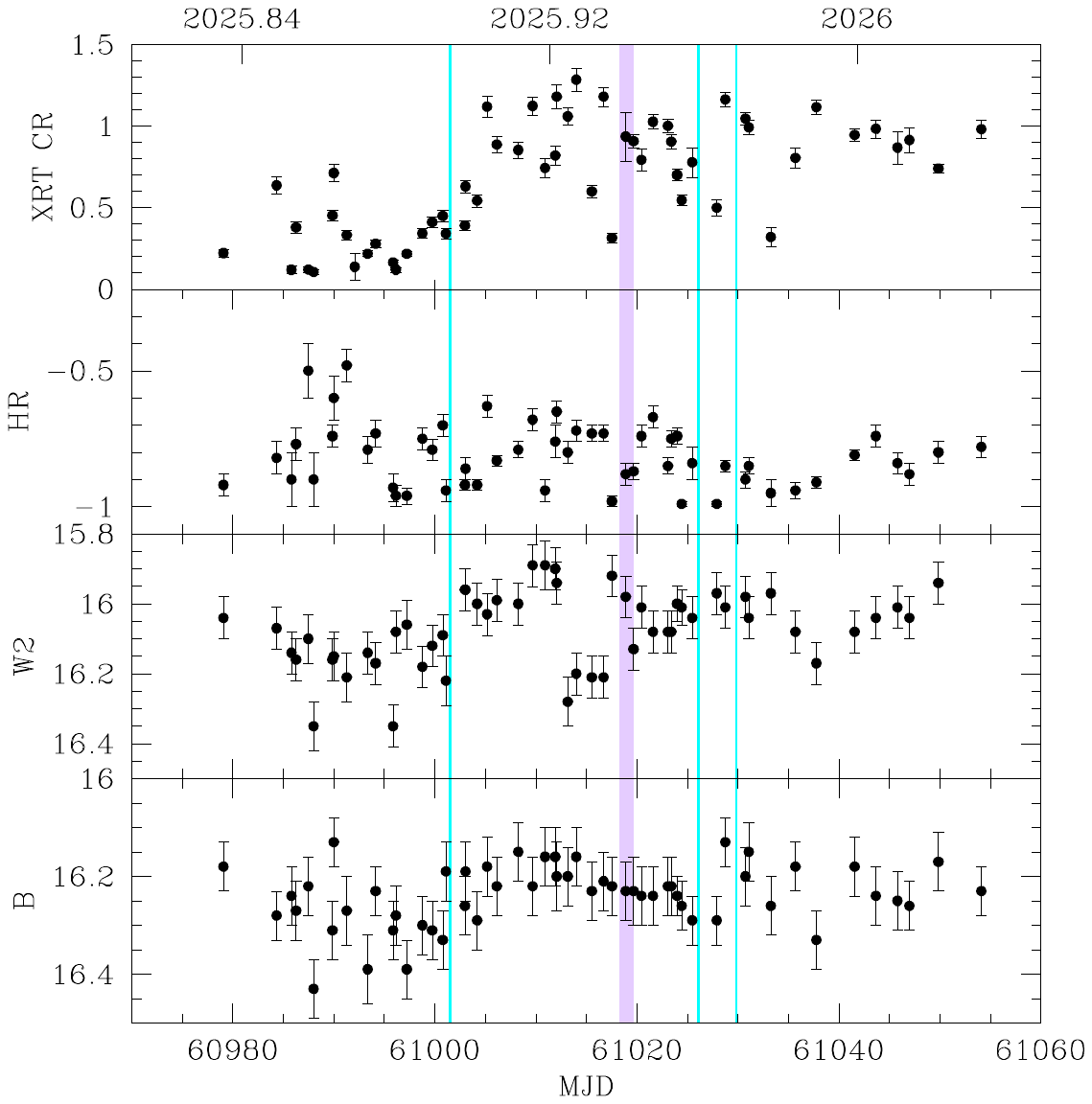}
    \caption{Outburst light curve of IC 3599 in the X-ray, UV $W2$, and $B$ band until 2026 January 14. $W2$ and $B$ are Vega magnitudes and are corrected for Galactic extinction. The cyan lines mark the times of the optical spectroscopic observations and the magenta box the time of the {\it XMM-Newton} observation. }
\label{outburst_lc}
\end{figure}

\subsection{\swift\ Results}

The new {\it Swift} monitoring observations allow us to address a number of key questions regarding  the X-ray spectral evolution and variability at high state for the first time. 

In a first step, we only focus on the simple hardness ratios (HR){\footnote{We define the hardness ratio by HR $= (H-S)/(H+S)$, where $S$ is the counts in the (0.3--1.0) keV band and $H$ in the (1.0--10.0) keV band.}, since that allows a comparison with low-state data. 
The hardness ratio was then estimated by applying the Bayesian Estimation of Hardness Ratios (BEHR) by \cite{park2006}}, and the $UVW2$ light curves of IC 3599 
are illustrated in Figure\,\ref{swift_lc}. This shows that the 2025  (0.3--2.0) keV outburst flux may have exceeded that of the previous 2010 outburst observations and also the 1990 outburst observations (not plotted).
The $UVW2$ light curve may suggest that the onset of the outburst might have already started around May 2025. Owing to the Sun constraint, we have no data regarding IC 3599's behavior between 2025 July 31 and October 30. In outburst, the spectrum of IC 3599 became much 
softer, with no HR values above $-0.5$ were measured{\footnote{Note that the HR of a typical AGN with $\Gamma=2$ is about 0.0.}}. 

Second, the light curve of the most recent 2025--2026 outburst is shown in Figure\,\ref{outburst_lc}. 
An important question is whether the spectrum evolves with the brightness level. However,
plotting HR versus CR, there is no particular trend visible. The X-ray spectrum
of IC 3599 is always very soft, but there is no trend of softness with brightness.

At the same time, there are clear differences in HR. Therefore, in the next step,
we selected the spectrum with the highest HR $= -0.5$, and a representative spectrum with a low HR $= -0.9$ (of December 28) and systematically (re)fit the spectra with a simple black-body representation. The absorption was fixed at the Galactic value,
$N_{\rm H}=1.35 \times 10^{20}$ cm$^{-2}$ \citep{HI4PI2016}. A single black-body model represents the HR $= -0.9$ spectrum well with $kT_{\rm BB}=(0.11\pm{0.03})$ keV, but it is not a good match to the other, harder spectrum. Therefore, in the next step, we have refit the HR $= -0.5$ spectrum ignoring the hard tail above 1 keV and fitting the soft spectral part only. This provides a consistent fit with 
$kT_{\rm BB}=(0.09\pm{0.01})$ keV, and the additional hard tail emerges at energies $\gtrsim 1$ keV, indicating an extra power-law component presumably from a faint corona.  

To measure the peak X-ray luminosity of IC 3599 during outburst, we  carried out a more detailed analysis of the spectral epoch with the highest count rate measured in WT mode, observed on 2025 December 18. 
The spectrum was first fit with a single black-body model with absorption fixed at the Galactic value. This provides a very good representation of the spectrum, except for a small excess above $\sim 1$ keV, and gives $kT_{\rm BB}=(0.12\pm0.01)$ keV and $\chi^2/\nu = 43.8/35$ (Appendix, Figure\,\ref{x-highstate-fit}). 
In the next step, in order to see how much excess absorption might be present along the line of sight, we have refit the spectrum with $N_{\rm H}$ as a free parameter. However, no excess absorption beyond the Galactic value is required.
We then used a more physical and broader disk black-body model to see if it can account for faint residuals beyond 1 keV in the black-body fit. However, the fit does not improve significantly. Next, a single power law was used. Despite a very soft photon index, $\Gamma_{\rm X} \approx 4$, the model gives a bad fit to the spectrum ($\chi^2_{\rm red} = 4.4$) with strong residuals (Fig.\,\ref{x-highstate-fit}).  
We take the black-body model fit (Fig.\,\ref{x-highstate-fit}) for further estimates. 
That model has an absorbed (unabsorbed) (0.3--10) keV flux of $(2.64\pm0.12)\times 10^{-14}$ W m$^{-2}$ ($3.0\times 10^{-14}$ W m$^{-2}$).
This flux corresponds to a (0.3--10) keV luminosity of $L = 3\times 10^{36}$ W.

Overall, the X-ray count rate during the outburst is highly variable and does not display a smooth rise to the maximum. The fastest factor-of-2 flux change happens
within 3.6 hr, and the highest amplitude of variability in subsequent observations is a factor of 5.3 within 36
hr.  As of 2026 January 20, the outburst has lasted more than 80 days. 
In order to search for rapid, high-amplitude flaring at short timescales that would potentially constrain the SMBH mass of IC 3599, we have analyzed the {\it XMM-Newton} pn light curve of IC 3599.

\section{XMM-Newton light curve}

\begin{figure}
\includegraphics[trim=0 120 0 150,clip,width=9.3cm]{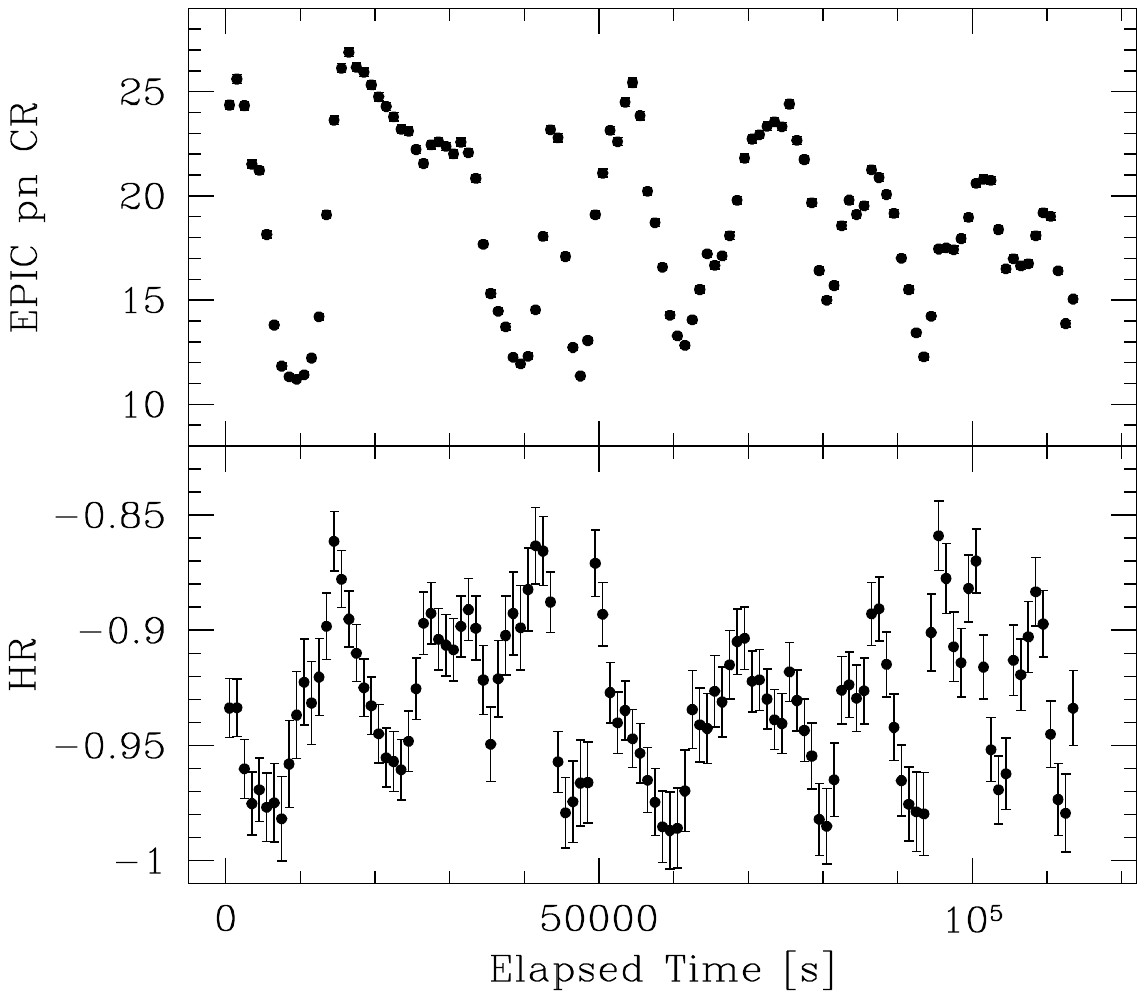}
    \caption{{\it XMM-Newton} EPIC pn light curve of IC 3599, revealing a strong oscillation pattern. The top panel shows the count rate and the lower panel the hardness ratio.}
\label{xmm_lc}
\end{figure}

We obtained a 120 ks {\it XMM-Newton} observation of IC 3599 on 2025 December 09--10 (observation ID 0973390701). A key goal of that observation is to search for absorption lines signaling outflows that could arise during the near-Eddington accretion phase at outburst. These results will be presented elsewhere (Grupe et al., in prep.).
In a search for variability on timescales more rapid than
can be traced by {\it Swift}, we inspected the {\it XMM-Newton} light curve 
that is presented here. 
Details on the data analysis are give in Section\,\ref{xmm_analysis} of the Appendix.
The CR and HR\footnote{For the pn data we defined the soft and hard energy ranges between (0.2--1.0) keV and (1.0--10.0) keV, respectively.} light curve (Fig.\,\ref{xmm_lc}) 
exhibits high-amplitude oscillatory variability that is
reminiscent of quasiperiodic oscillations.
A Lomb-Scargle periodogram 
of the count-rate light curve 
shows the strongest signal at $3.76\times 10^{-5}$ Hz (or a period of 26,600 s, or 7.4 hr) as described in Appendix\,\ref{lsp}.  However, given that the length of the observation only covers four such periods and the light curve clearly shows more substructure,
this result needs verification in deeper follow-up observations.  
 The fastest change in flux is observed at the beginning of the {\it XMM-Newton} observation with a rise by a factor of 2.5 within 1.75 hr. 
 The HR light curve closely follows the CR light curve. 
We have also inspected the MOS light curves, and they show the same variability pattern. 
Further details of the {\it XMM-Newton} observation will be discussed in future work. 

\section{Optical Spectroscopy }

In order to search for an emission-line response to the giant-amplitude outburst, we have carried out optical spectroscopy \citep{Komossa2025} of IC 3599 at Lick Observatory (on 2025 November 22), at Xinglong Observatory (on 2025 December 20), and at the Caucasian Mountain Observatory (CMO, on December 16). Data acquisition and reduction were carried out in a standard way, as described in  Appendix\,\ref{opt_obs}. As a last step before emission-line analysis, all three spectra were homogeneously corrected for a Galactic extinction of $E_{B-V}=0.015$ mag, using the Galactic reddening law of \cite{cardelli1989}, assuming a reddening coefficient $R_V=3.1$. 

\begin{figure}
\includegraphics[trim=10 40 60 10,clip,width=6.3cm, angle=270]{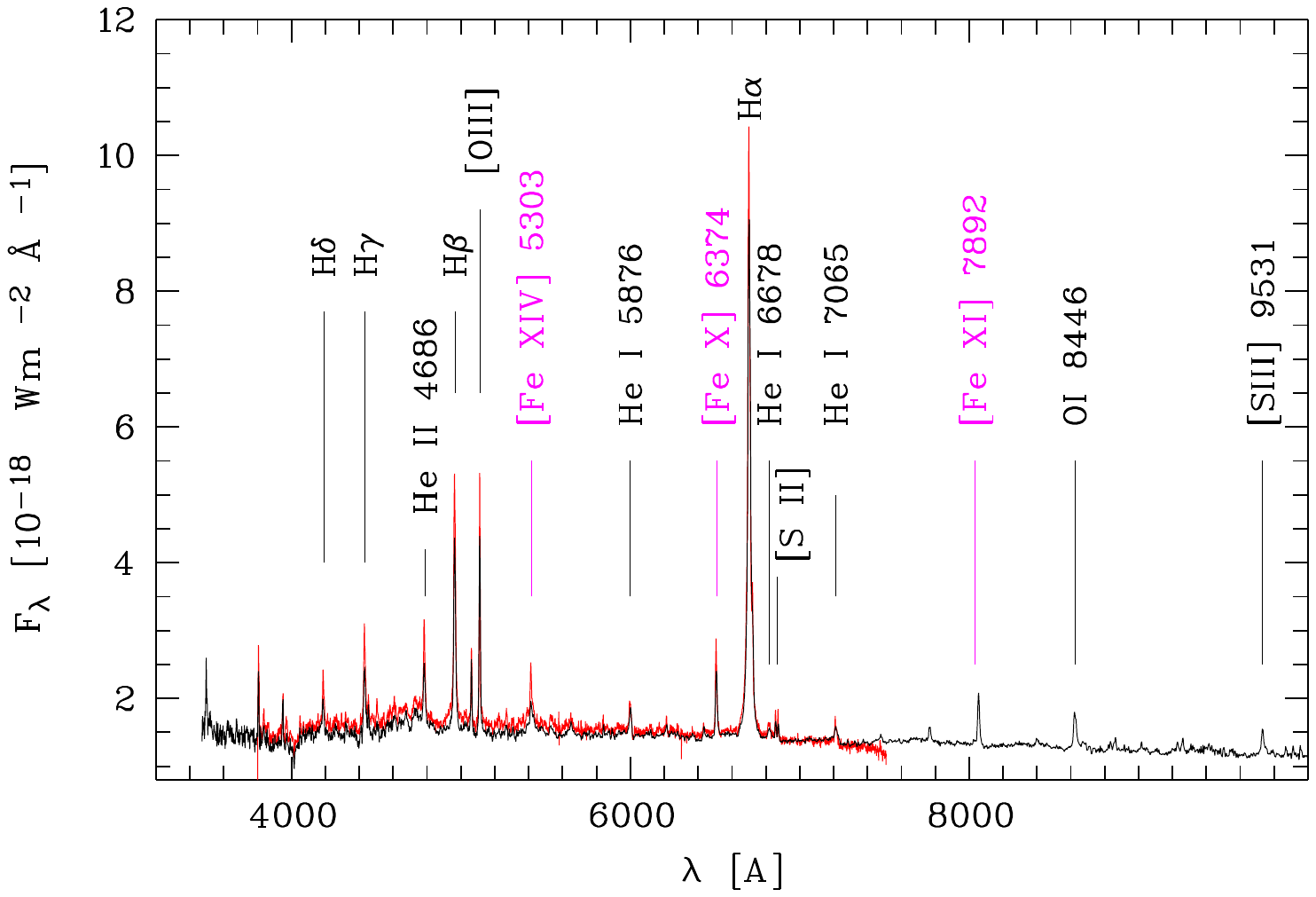}
    \caption{Optical spectrum of IC 3599 taken at Lick Observatory on November 22. The CMO spectrum from December 16 is plotted in red. 
    The brightest coronal emission lines are labeled in magenta. Other bright emission lines
    are labeled in black. \label{opt_spec}}
\label{optspec}
\end{figure}

The Xinglong spectrum clearly shows increased fluxes in the Balmer lines 
and displays several coronal lines. 
The relative strengths of the strong transitions of [Fe~X] $\lambda$6374, Balmer lines, and [O~III] $\lambda 5007$ agree with the results of the other spectra. 
Owing to the higher signal-to-noise ratio (S/N) of the Lick spectrum, and the larger number of detected transitions, further emission-line measurements are carried out on that spectrum. 

The Lick and CMO  spectra are displayed in Figure\,\ref{optspec}. 
The new spectra clearly exhibit a wealth of emission lines, many not present in the low-state spectra taken in the 1990s and with SDSS, while others have dramatically brightened. Prominent emission features include the Balmer lines, [O~III] $\lambda$5007, 
coronal Fe lines such as [Fe~XIV] $\lambda 5305$,  [Fe~VII] $\lambda 6087$, [Fe~X] $\lambda 6374$, and [Fe~XI] $\lambda 7892$, and the He~I transition at 6678~\AA.

For further analysis, we fitted the bright emission lines with
Gaussian profiles, and measured line widths and fluxes. Note that the BLR and CLR emission discussed below is so compact that it is always within the slit. 
At a scale of 0.4 kpc/arcsec, a significant part of the NLR is additionally within the slit (see also the note below). 
The primary analysis was done on the Lick spectrum that overall agrees well with the CMO and Xinglong spectra (Fig.\,\ref{opt_spec}). 

The final instrumental resolution owing to spectrograph, slit, and weather conditions was determined based on night-sky lines in the two-dimensional (2D) spectra, and resulted in a full width at half-maximum intensity (FWHM$_{\rm inst}$) = $300 \pm 20$ km s$^{-1}$ in the blue and $430 \pm 40$ km $^{-1}$ in the red. The measured FWHM values were then corrected for instrumental resolution (Table\,\ref{opt_lines}), following the standard quadrature equation FWHM$_{\rm corr} = \sqrt{{\rm FWHM}_{\rm obs}^2 - {\rm FWHM}_{\rm inst}^2}$. The uncertainties reported in the emission-line measurements were determined as described in Appendix\,\ref{opt_obs}. 

A very striking difference with the low-state spectra of IC 3599 is the dramatic increase of the high-ionization coronal lines  [Fe~X]--[Fe~XIV], and the shift in ionization states: [Fe~X]--[Fe~XIV] are undetected in the low-state spectra \citep[Figs. 6 and 7 of][]{KomossaBade1999}, but are very prominent in the 2025 outburst spectra (Fig. 4). The low-state spectra instead show several transitions of [Fe~VII], a much lower degree of ionization. 
Furthermore, there is evidence that 
the width of H$\beta$ has increased. In the low state, FWHM(H$\beta$) = 480 km s$^{-1}$ \citep[corrected for instrumental resolution;][]{KomossaBade1999}, compared to $820 \pm 60$ km s$^{-1}$  (corrected for instrumental resolution) in the high state. 

Moreover, the flux ratio of H$\beta$ to [O~III] changed dramatically.
If one compares Figures 6 and 7 of \citet{KomossaBade1999} with our Figure 4, one can see that the spectrum is dominated by [O~III] in the low state, but by H$\beta$ in the high state (and the same statement holds for the 1991 high-state spectrum of \citealt{Brandt1995}). Of course,  small variations in [O~III] strength may be expected from a contribution of the NLR owing to the slit covering different parts of the outer NLR and varying seeing. However, our inspection of the 2D spectra in the cross-disperson direction shows  that the spatial profiles of H$\beta$ and [O~III] $\lambda$5007 are very compact and similar, and decline rapidly with radial distance, so we expect them to be similarly affected by the seeing. 
Specifically, fitting a Gaussian to the cross-dispersion distribution of [O~III] $\lambda$5007 and H$\beta$, we find that the intensity of [O~III] and H$\beta$ is at 1/10 of its central value at $2.9''$ and $3.1''$, respectively. 
In the low-state spectra of the 1990s \citep{Grupe1995a, KomossaBade1999} and the 2005 SDSS spectrum (Komossa et al. 2026, in prep.), the narrow-line flux ratio H$\beta$/[O~III] was $\sim 0.30$. Now, in the high state, we measure a ratio of the 
entire H$\beta$ line to [O~III] of $1.4\pm0.1$, corresponding to an increase of a factor $>4.7$.

\begin{table}
 \caption{Bright lines in the Lick spectrum of IC 3599.
 \label{opt_lines}}
\centering
  \begin{tabular}{lcccc}
  \hline
\hline
Line & $\lambda_0$ [\AA] & FWHM$^1$ & Flux$^2$   \\
\hline
H$\delta$ & 4102 &  1040\plm125 & 8.4\plm1.5   \\
H$\gamma$ & 4341 &  900\plm80 & 15.9\plm1.6   \\
He II     & 4686 &  800\plm60 & 12.9\plm0.8 \\
H$\beta$ & 4861 & 820\plm60 & 43.9\plm2.1 \\
{[O III]} & 4959 & 400\plm50 & 9.5\plm1.6  \\
{[O III]}   & 5007 & 480\plm35 & 30.3\plm1.7 \\
{[Fe XIV}] & 5303 & 1800\plm160$^{4}$ & 13.5\plm1.5  \\
He I     & 5876 & 560\plm60 & 6.6\plm0.6  \\
{[O I] }    & 6300 & 480\plm80  & 2.7\plm0.6  \\
{[O I]+[Fe X]} & 6374 & 400\plm40 & 14.0\plm0.7   \\
H$\alpha$$^5$   & 6563 & 780\plm75 & 159.8\plm8.2 \\
{[N II]}     & 6584  & 400\plm80 & 21.3\plm6.3  \\
He I        & 6678  & 930\plm95 & 3.9\plm0.4  \\
{[S II]}     & 6717   &   430$^3$ & 3.1\plm0.3  \\
{[S II]}     & 6731   &  380$^3$ & 2.5\plm0.3 &  \\
He I        & 7065 & 900\plm125 & 4.9\plm0.8  \\
{[Fe XI] }  & 7892 & 360\plm50 & 11.3\plm0.7 \\
{O I}     & 8446 & 730\plm70 & 12.5\plm1.0 \\
{[S III] }  & 9531 & 440\plm50 & 7.2\plm0.8  \\
\hline
\hline
\end{tabular}
$^1$FWHM given in units of km~s$^{-1}$ corrected for instrumental resolution.
$^2$Line fluxes are given in units of $10^{-18} $ W m$^{-2}$ ($10^{-15}$ cgs). 
$^3$Unresolved.  
$^4$The measured FWHM and flux are upper limits owing to the possible contribution of other Fe transitions (Komossa et al. 2026, in prep.).
$^5$Blended with [N II] $\lambda6548$. Given the theoretical line ratio of [N~II] $\lambda6584$/[N~II] $\lambda6548$ = 3, the impact on the H$\alpha$ measurement is small.  
\end{table}

\section{Discussion and Conclusions}

\subsection{X-ray Peak Luminosity and Eddington Ratio \label{disc_luminosity}}

Based on the spectral fit to the {\it Swift} high state, we obtained a (0.3--10) keV X-ray luminosity of $3\times 10^{36}$ W 
(= $3\times 10^{43}$ erg s$^{-1}$). Using the (0.2--2.0) keV luminosity of $4\times 10^{36}$ W and the bolometric correction given by \cite{Grupe2010}, the bolometric luminosity is $4.6\times 10^{37}$ W.

The SMBH mass of IC 3599 cannot directly be estimated from the Balmer lines, since observed Balmer lines are dominated by CLR emission, and the  broad-base component is not reliably measured and variable as well. Instead, we use the SMBH mass estimated from host-galaxy scaling relations, of $2.2 \times 10^6~M_{\odot}$  \citep{Grupe2015}. Assuming this black hole mass, the Eddington luminosity is $3\times 10^{37}$ W, which implies that $L/L_{\rm Edd}$ of IC 3599 at peak is about unity\footnote{During the low state the Eddington ratio was $L/L_{\rm Edd} \approx0.01$.}. 

\subsection{Optical Spectra}

The 1991 outburst optical spectrum, in addition to bright coronal emission lines, showed unresolved Balmer lines
\citep{Brandt1995}. The low-state spectrum  taken $\sim 6$ yr later 
had a broad component detected in H$\alpha$ with a FWHM of 1090 \kms \citep{KomossaBade1999}, that would formally make IC 3599 a narrow-line Seyfert 1.9 (NLS1.9), but without Fe II emission.
The new 2025 outburst spectrum shows classical long-lived NLR emission lines, and substantially brightened high-ionization lines from a CLR, powered by the bright outburst. The bulk of the Balmer lines are narrow with a FWHM of $\sim 800$--900 \kms and appear to be associated with the CLR. 
The transition of He~I 6678 is detected for the first time in IC 3599, and demonstrates the broad range of ionization states triggered in response to the outburst. 
Monitored over time, the strong emission-line response of IC 3599 provides us with a unique case of (BLR and) CLR reverberation mapping using a wealth of emission lines \citep{KomossaBade1999, Komossa2008}, as the light echo travels through the gas-rich core region of the galaxy
(see \citealt{Smith2025} for a recent application making use of variable [NeV] in a quasar). 
Further implications of the optical spectrum and ongoing optical monitoring will be discussed in future work (Komossa et al., in prep.).

\subsection{Outburst Models}

The high amplitude of variability of IC 3599 and its strong emission-line response makes it an exceptional case of an optical changing-look AGN. A variety of theoretical models for changing-look AGN have been explored in recent years aimed at explaining fast changes in the accretion rate, much faster than the viscous timescale \citep[e.g.,][]{Stern2018, NodaDone2018, Lawrence2018, DexterBegelman2019, Sniegowska2020, Laha2022, Wang2024, LiCao2025}. 
Changes on timescales of decades can be well explained by a radiation-pressure instability \citep{Lightman1974}, whereas 
changing-look events as rapid as months as observed in, for example, NGC 1566 \citep{Alloin1986, Parker2019, Oknyansky2020, Ochmann2024} require a different explanation. 

We previously favored an accretion-disk instability model of IC 3599 based on the
radiation-pressure instability \citep{Belloni1997, saxton2015}. That model \citep[see Eq. (1) of][] {Grupe2015} 
still explains the new outburst well, if we assume local changes in the long-lived accretion disk that impact the onset and timing of the next cycle such that recurrent outbursts are not strictly periodic.
At $\Delta t_2 = 15.7$ yr
and all other parameters as before {{\citep[$M_{\rm Edd}=1$, $R_0=3\,R_g$, $\alpha=0.1$;][]{Grupe2015}}, we obtain a truncation radius of $R_{\rm trunc} = (6.2$--38)\,$R_g$ for a range in $M_{\rm BH} = (10^6$--$10^7)~M_{\odot}$. Using the black hole mass of $2\times 10^6~ M_{\odot}$ measured from host scaling relations, the truncation radius is 20~$R_g$. 
This estimate assumes Eddington accretion at peak, in excellent agreement with the actual {\it Swift} observations 
(Sec.\,\ref{disc_luminosity}). 
We also note that decades-old and more recent radiation-pressure instability models \citep[e.g.,][]{sniegowska2023} show that the flare morphology is model-dependent, being sensitive to, for instance, the disk-corona structure, magnetic field, and outer disk radius. The observations of IC 3599 are therefore consistent with a radiation-pressure-instability interpretation, but they do not uniquely distinguish between different implementations of that scenario.

Alternative scenarios that produce recurrent flares at constant time intervals $\Delta t$ can be ruled out, since $\Delta t_1 = 19.5$ yr and $\Delta t_2 = 15.7$ yr. Such models include variants of an orbiting star undergoing repeat tidal stripping events, and of an orbiting compact object (such as a second SMBH) disturbing the inner accretion disk repeatedly. A shrinking orbital timescale and/or an elliptical/precessing orbit is still possible in some TDE repeat tidal stripping models, as the star loses energy during each interaction, and irregular episodes of brightening and dipping in TDE
binary SMBH scenarios \citep{Liu2009} formally remain possibilities as well, even though the similar outburst peak fluxes of IC 3599 then remain unexplained. Strong coronal-line responses (``extreme coronal-line emitters'') have also
been observed in {\em{quiescent}} galaxies and these cases have been interpreted as TDEs \citep[e.g.,][]{Komossa2008, Wang2012, Callow2025}. 
However, the fact that IC 3599 is a long-lived AGN with a long-lived accretion disk and NLR, rather calls for accretion-disk-related outburst models for IC 3599.
In a binary SMBH scenario, it is possible that the orbital timescale is shrinking if the binary is in the gravitational-wave-driven inspiraling regime already, but in order to change as rapidly as observed, a coalescence would be imminent, which we consider highly unlikely. Further, a highly elliptical orbit was needed, if one of the SMBHs would be impacting the other's inner disk.

The short-term oscillatory pattern of the {\it XMM-Newton} light curve that is consistent with a period of $\sim 7.4$ hr requires special attention. 
Reliably identified X-ray periodicities on timescales of minutes to days are rare in AGN \citep{Vaughan2005, Gierlinski2008, Middleton2011, Masterson2025}. There are several exceptions, and it is interesting to compare them with the case of IC 3599. 
First, quasiperiodic eruptions (QPEs) have been detected in a few AGN \citep[e.g.,][]{Giustini2020, Miniutti2023, Webbe2023}. These sources show exceptionally soft X-ray spectra and thus share similarities with the outburst spectra of IC 3599 \citep[][our Fig. \ref{x-highstate-fit}]{Grupe2015}. However, QPEs are typically characterized by sharp flares lasting hours, with longer episodes of inactivity in between, significantly different from the light curve of IC 3599 (Fig. \ref{xmm_lc}). The physics behind QPEs remains uncertain, and several different scenarios have been proposed, including disk instabilities, disk precession, or the presence of orbiting compact objects interacting with the disk. 
Second, quasiperiodic oscillations (QPOs) with a period of 1 hr have been identified in an {\it XMM-Newton} observation of the NLS1 galaxy REJ1034+396 \citep{Gierlinski2008}. The physical driving mechanism remained unknown.  Overall, the light curve of IC 3599 appears very similar, albeit with a longer candidate period of 7.4 hr (or possibly 4.8 hr; see Appendix D). 
Adopting a SMBH mass of $2 \times 10^6~ M_{\odot}$ (see above), that period (4.8--7.4 hr) would correspond to a Keplerian orbit with a semimajor axis  of $a = (0.8$--1.1) AU (or 40--60~$R_g$). 
However, since the {\it XMM-Newton} light curve only covers a few periods of the strongest peak at a period of 26,600 s
in the Lomb-Scargle periodogram, we conclude that longer coverage is required to confirm or reject the candidate period.
Alternatively, irregular but high-amplitude variability in the (near- or super-)Eddington regime remains a possibility.
The {\it XMM-Newton} data and their implications will be discussed in more detail in future work.

In summary, IC 3599 is among the most extreme and exceptional changing-look AGN known with an X-ray light curve that dates back 3.5 decades.
Its dramatic, supersoft X-ray outbursts, recurrent within decades, power a strong optical emission-line response with a broad range of low- and high-ionization transitions discovered in optical spectra. Accreting at the Eddington limit at the peak of the outburst, and with an unusual QPO-like pattern in its {\it XMM-Newton} outburst light curve, IC 3599 provides us with unique insights into accretion physics under extreme conditions.

\begin{acknowledgments} 
We thank   
Brad Cenko for approving our requests to observe IC 3599 and the \swift\ Science Operations team for executing the observations. 
We also thank Erik Kuulkers for granting our \xmm\ DDT XMM-Newton request, Lucia Ballo for excellent communications and help during the observation planning process, and the XMM-Newton Guest Observer facility at the National Aeronautics and Space Administration (NASA) Goddard Space Flight Center for their excellent documentation of the XMM-Newton data analysis. 
We are grateful to the anonymous referee for their comments. 

This research made use of the NASA/IPAC Extragalactic Database (NED) which is operated by the Jet Propulsion Laboratory, Caltech, under contract with NASA. 
This work used data supplied by the UK Swift Science Data Centre at the University of Leicester \citep{evans2007}. This research is partly based on observations obtained with {\it XMM-Newton}, an ESA science mission
with instruments and contributions directly funded by
ESA Member States and NASA.

S.K. is grateful to the CAS President's International Fellowship Initiative, and NAOC Beijing for their hospitality. 
V.O. acknowledges financial support  by Israeli Science Foundation grants 2398/19 and 1650/23.
A.V.F.'s research group at U.C. Berkeley was supported by the Christopher R. Redlich Fund, Gary and Cynthia Bengier, Clark and Sharon Winslow, Alan Eustace and Kathy Kwan (W.Z. is a Bengier-Winslow-Eustace Specialist in Astronomy), Timothy and Melissa Draper, Briggs and Kathleen Wood, Ellyn and Alan Seelenfreund (T.G.B. is Draper-Wood-Seelenfreund Specialist in Astronomy), and numerous other donors.

A major upgrade of the Kast spectrograph on the Shane 3 m
telescope at Lick Observatory, led by Brad Holden, was made possible through
generous gifts from the Heising-Simons Foundation, William and Marina Kast,
and the University of California Observatories. Research at Lick Observatory is
partially supported by a generous gift from Google.

\end{acknowledgments} 

\vspace{5mm}
\facilities{NASA Neil Gehrels Swift Observatory (XRT and UVOT), XMM-Newton, UCO Lick Observatory Shane 3 m telescope, Xinglong Station 2.16 m telescope, and Caucasian Mountain Observatory 2.5 m telescope}

\software{HEASoft (\url{https://heasarc.gsfc.nasa.gov/docs/software/heasoft/}) with XSPEC \citep{arnaud1996}, ESO-MIDAS (\url{https://www.eso.org/sci/software/esomidas/}), 
the R programming language (\url{https://www.r-project.org/}), 
 SuperMongo  (\url{https://www.astro.princeton.edu/~rhl/sm/}), the Image Reduction and Analysis Facility \citep[IRAF, ][]{tody1986},
 and the 
  XRT Data Analysis Software (XRTDAS) developed under the responsibility
  of the ASI Science Data Center (ASDC), Italy.
}

\section{Data Availability}
Data are available or will become available in public archives after proprietary periods. 
The {\it Swift} data of our project are available in the {\it Swift} archive at \url{https://swift.gsfc.nasa.gov/archive/}.
Our {\it XMM-Newton} data are in the archive at \url{https://www.cosmos.esa.int/web/xmm-newton/xsa}. 
The CMO data are available upon reasonable request. The Xinglong data will appear in the Xinglong archive that is being set up. 
The Lick data can be accessed in the archive at \url{https://mthamilton.ucolick.org/data/}.

\newpage

\appendix

\section{Details on the Swift Observations of the Outburst}
All \swift\ observations were performed under the target ID 10375 starting with sequence number 135. The majority of our observations reported here were done in a daily monitoring campaign. This cadence was later relaxed to a two- and three-day monitoring. 
Most of the XRT and UVOT observations had exposure times of 800--1000~s each.

\section{High-state Swift X-ray Spectral Fit}

Here we show the results of the high-state X-ray spectral fit. The outburst spectra of IC 3599 are very soft, with only a faint hard X-ray spectral component
present at some states, presumably representing a faint corona. The highest-state spectrum taken in the WT mode of 2025 December 19 is illustrated in Figure\,\ref{x-highstate-fit}. It is well described by a black-body emission model. 

\begin{figure}
\centering
\includegraphics[width=8.5cm]{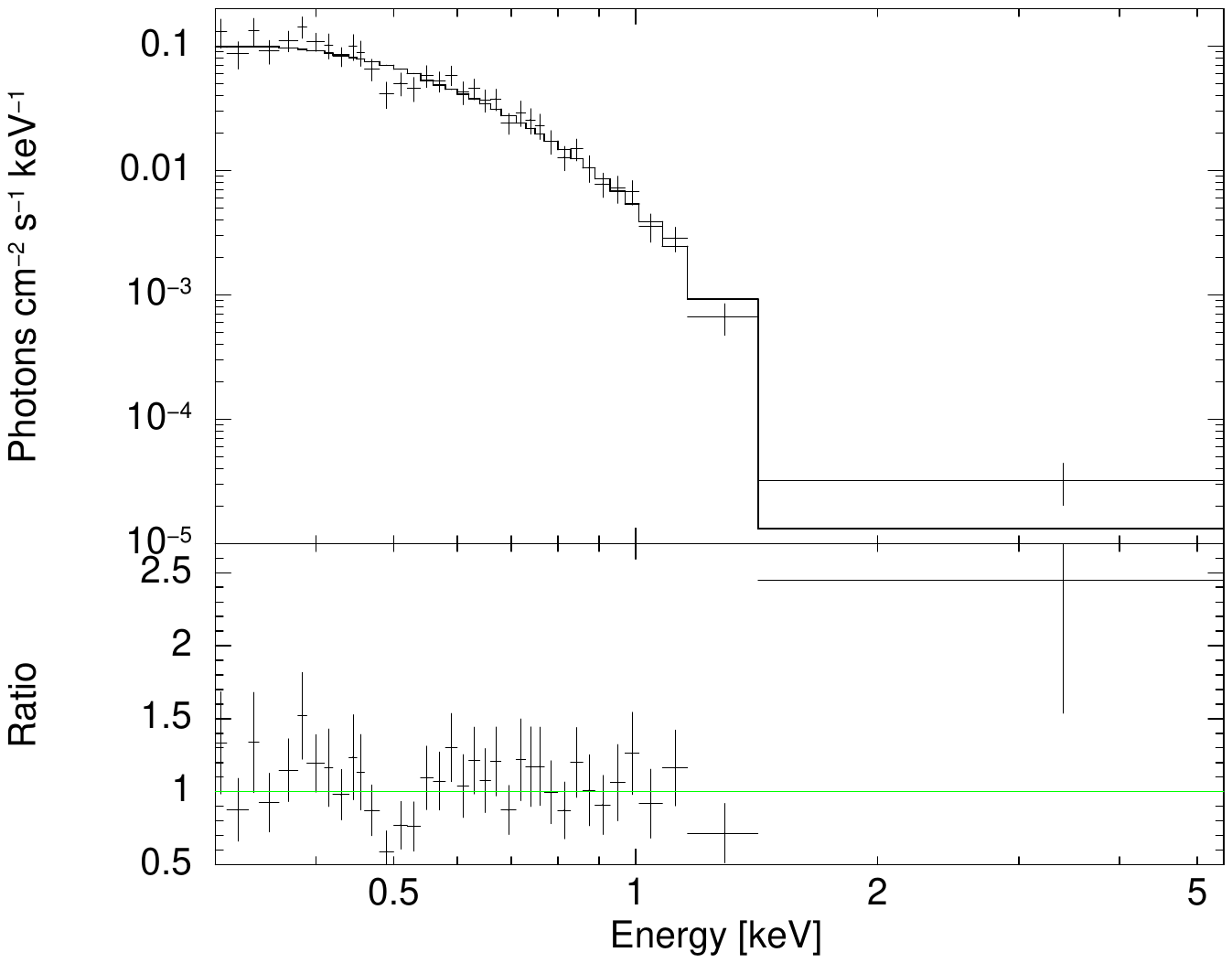}
\includegraphics[width=8.5cm]{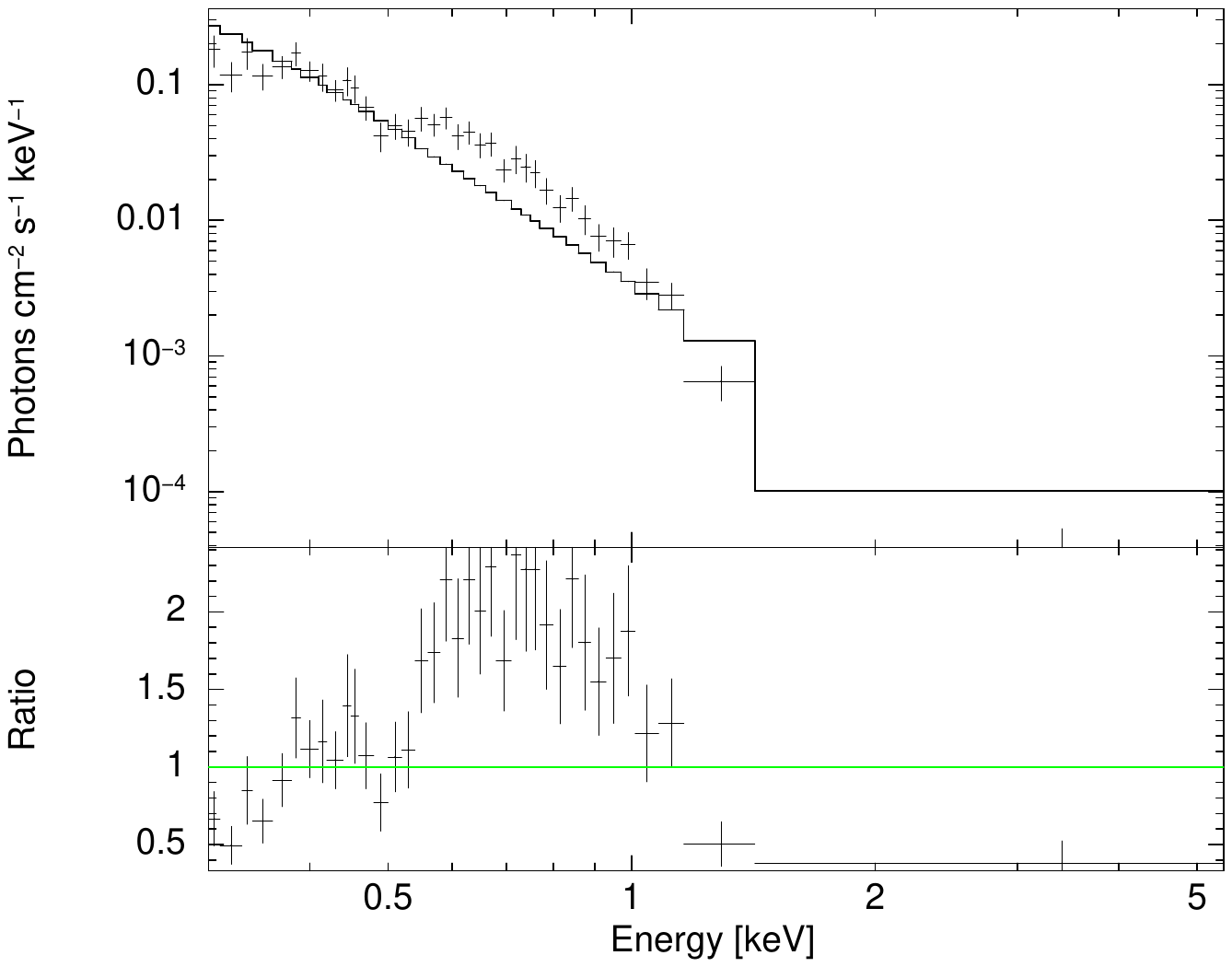}
    \caption{
   Left:}} High-state X-ray spectrum of IC 3599 of 2025 December 19 fit with a single black-body model  {\bf{and absorption fixed at the Galactic value (Sec. 2.3)}}. The lower panel shows the fit residuals. The spectrum is exceptionally soft.   {\bf{Right: The same spectrum fit with a single power-law model (Sec. 2.3) that does not provide a successful description. 
    }
\label{x-highstate-fit}
\end{figure}

\section{XMM-Newton Data Analysis \label{xmm_analysis}}

After the {\it Swift} detection of the X-ray outburst of IC 3599 we 
requested a director's discretionary time observation with {\it XMM-Newton}, which was executed on 2025 December 09/10 for a total of 120 ks. Because here we only focus on the EPIC pn light curve, we limit the data-analysis description to the pn data light curve only. 

Data were analyzed with the newest version of the XMM-SAS, 22.1.0. The observation was performed in small window mode to minimize pile-up. 
We first inspected the data for periods of high background flares and found that except for the last 4 ks of the observation, the rest of the exposure did not exhibit background flares. 
Owing to the high count rates with up to almost 30 counts s$^{-1}$, we limited the analysis to single events only (pattern=0). Source and background counts were selected from circular regions with a radius of $30''$, and source and background light curves were created with {\it evselect} with a binning of 1000 s. The source light curve was then background subtracted with the {\it epiclccorr} tool. 

In order to create the hardness ratio light curve we separately made soft and hard source and background light curves in the (0.2--1.0) keV and (1.0--10) keV ranges, respectively. Owing to the significant number of counts in each channel, the hardness ratios were directly calculated as HR $=(H-S)(H+S)$.

\section{Lomb-Scargle Periodogram \label{lsp}}

For a first evaluation of possible periodicity in the {\it XMM-Newton} EPIC pn light curve of IC 3599 (Fig.\,\ref{xmm_lc}), we have computed the Lomb-Scargle periodogram \citep{lomb1976, scargle1982}. 
In this approach, trial sinusoidal light‑curve models are fitted to the data. The program determines how well at a given frequency the variance seen in the data can be explained by a sinusoidal behavior given the trial frequency. A normalized power of $P(\nu)=1$ corresponds to a perfect sinusoidal light curve. 

The analysis was performed in R \citep{R-core, crawley2012}
and shows the main peak at a frequency of $3.76\times 10^{-5}$ Hz, corresponding to a period of 26600~s (7.4 hr) with a normalized power of $P(\nu)=0.27$. This corresponds to a $4.5\sigma$ detection with a confidence level of 99.9992\% within this approach.
The second-strongest signal is found at a frequency of $5.75\times 10^{-5}$ Hz corresponding to a period of 17360~s (4.8~hr) with a normalized power $P(\nu)=0.20$. This is a $3.3\sigma$ detection with a confidence level of 99.95\%.

\begin{figure}
\centering
\includegraphics[width=8.3cm]{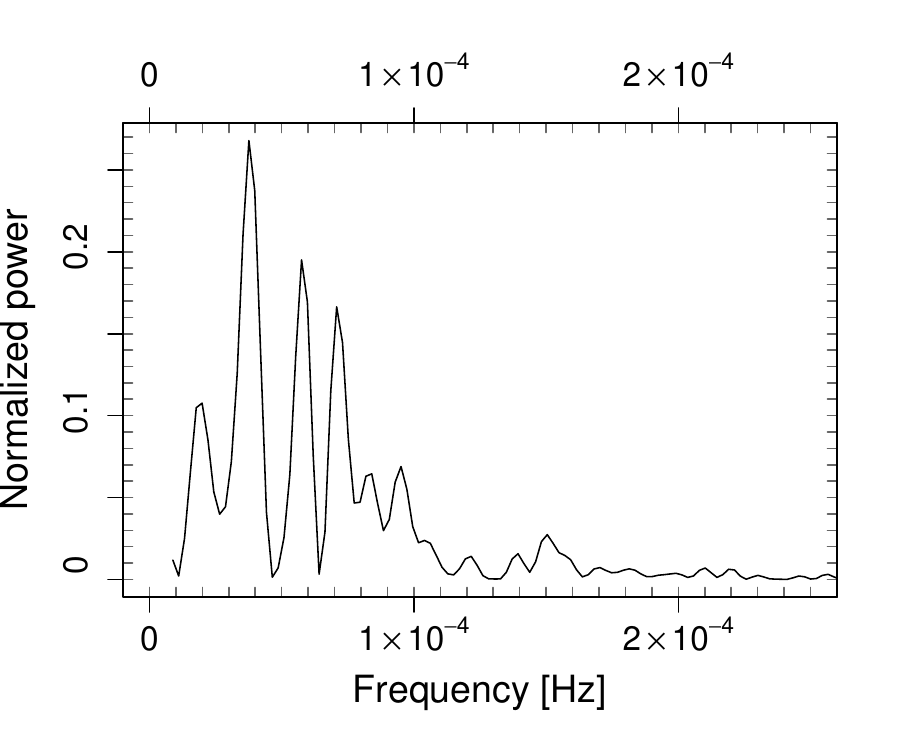}
    \caption{Lomb-Scargle periodogram of the {\it XMM-Newton} EPIC pn light curve shown in Figure\,\ref{xmm_lc}.  }
\label{pn_lsp}
\end{figure}

\section{Optical Spectroscopy \label{opt_obs} }

\subsection{Lick Spectroscopy} 
An optical spectrum was acquired with the Kast double spectrograph \citep{miller1994}
 mounted on the 3~m Shane telescope at Lick Observatory, California,
 on 2025 November 22. 
 The observation utilized the $2''$-wide slit, the D57 dichroic, the 600/4310 grism, and the 300/7500 grating.  
This instrument configuration has a combined wavelength range of $\sim 3600$--10,700~\AA, and a spectral resolving power of $R \approx 800$.  To minimize slit losses caused by atmospheric dispersion \citep{filippenko1982}, the slit was oriented at or near the parallactic angle. 
The data were reduced following standard techniques for CCD processing and spectrum extraction \citep{silverman2012}
 utilizing standard IRAF \citep{tody1986}
 routines and custom Python and IDL codes\footnote{https://github.com/ishivvers/TheKastShiv}.  
Low-order polynomial fits to comparison-lamp spectra were used to calibrate the wavelength scale, and small adjustments derived from night-sky emission lines in the target frames were applied. 
The spectra were flux calibrated using observations of appropriate spectrophotometric standard stars observed on the same night, at similar airmasses, and with an identical instrument configuration. Telluric features were also removed through comparison with these spectra. 

The final instrumental resolution 
was determined based on night-sky lines in the 2D spectrum by fitting the lines with a Gaussian profile. The source's emission-line fluxes and FWHM values were determined by applying the {\it deblend} command within MIDAS assuming a local continuum. {\it Deblend} uses a damped least-squares fitting (Levenberg–Marquardt algorithm) of the lines. The uncertainties in the line measurements are determined from the covariance matrix. 
Final uncertainties in the line width were determined from the uncertainties from the measurements of the object's emission-line width and the measurements of the night-sky lines. In the blue part of the spectrum, the resolution was measured  with only one bright night-sky line; however, in the red the resolution was determined from the mean value measured from several night-sky lines.

\subsection{Caucasian Mountain  Observatory}

Another spectrum was taken with the 2.5~m telescope at the Caucasian Mountain Observatory (CMO) which is part of the Sternberg Astronomical Institute of the M. V. Lomonosov Moscow State University on 2025 December 16 in the B (3600--5770~\AA) and R (5670--7460~\AA) channels of the Transient Double-beam Spectrograph \citep{potanin2020}
 with the $1''$ slit, providing a resolving power of $R \approx 1000$ and 2400 in the B and R channels, respectively. The position angle of the slit was at 0$^\circ$ instead of the parallactic angle of $-35^\circ$, but atmospheric dispersion should be low owing to a high altitude of 68$^\circ$ (airmass 1.08). The exposure times in both channels were 900 s. Standard optical data reduction was applied with a custom Python code \citep{potanin2020}, including dark subtraction, 2D wavelength calibration with comparison-lamp spectra, and flat-field correction. Night-sky lines are used to adjust wavelength shifts, which gives an accuracy better than 0.5{~\AA} and 0.1~{\AA} in the B and R channels, respectively. Spectra were extracted by a simple summation of counts above the background in an aperture of $6''$. The flux calibration and telluric correction were performed with a spectrum of the standard star
 HD~63160 taken just before the observation of IC 3599 at an airmass of 1.2. The airmass difference between IC 3599 and HD~63160 was corrected by applying a mean atmospheric transmission curve at CMO. Owing to  slit losses on the narrow slit, the absolute flux may be incorrect.

\subsection{Xinglong Spectroscopy}

A long-slit spectrum of IC~3599 was obtained with the 2.16~m telescope \citep{fan2016}
at the Xinglong
Observatory of the National Astronomical Observatories, Chinese Academy of Sciences (NAOC), on 
2025 December 20 (UTC). An earlier spectrum taken on 2025 December 5 was not further analyzed owing to the strong wind, effectively resulting in a seeing of $\sim5^{\prime\prime}$. 
The Beijing Faint Object Spectrograph and Camera was used in the observation.
The instrumental setup with
a spectral resolution of $\sim$10~\AA\ and a wavelength coverage of 3600--8700~\AA\ is provided
by the use of the G4 grism, along with a long slit of width 1.85$^{\prime\prime}$ oriented in the North–South direction.
In order to enhance the signal-to-noise ratio and to remove cosmic rays easily,
the object was observed successively twice with an exposure time of 1800~s for each frame.
The Kitt Peak National Observatory standard star HD~109995 was observed for flux calibration after the exposure of IC 3599.

The raw frames were reduced by the IRAF package \citep{tody1986}
by following standard procedures,
including bias subtraction and flat-field correction. The 1D spectrum was then extracted
after combining the two exposures.
Wavelength and flux calibrations were performed by the spectra of the iron–argon comparison lamps and
the standard star, respectively.
The accuracy of the wavelength calibration is better than 2 \AA.
The telluric A band (7600--7630 \AA) and B band ($\sim 6860$ \AA) absorption caused by atmospheric O$_2$ molecules
were  removed from the calibrated 1D spectrum, based on the standard-star spectrum.
A Galactic extinction law with $R_V = 3.1$ \citep{cardelli1989} was adopted for the extinction correction. The spectrum is shown in Figure\,\ref{Xinglong}. 

\begin{figure}
\centering
\includegraphics[width=8.5cm, angle=270]{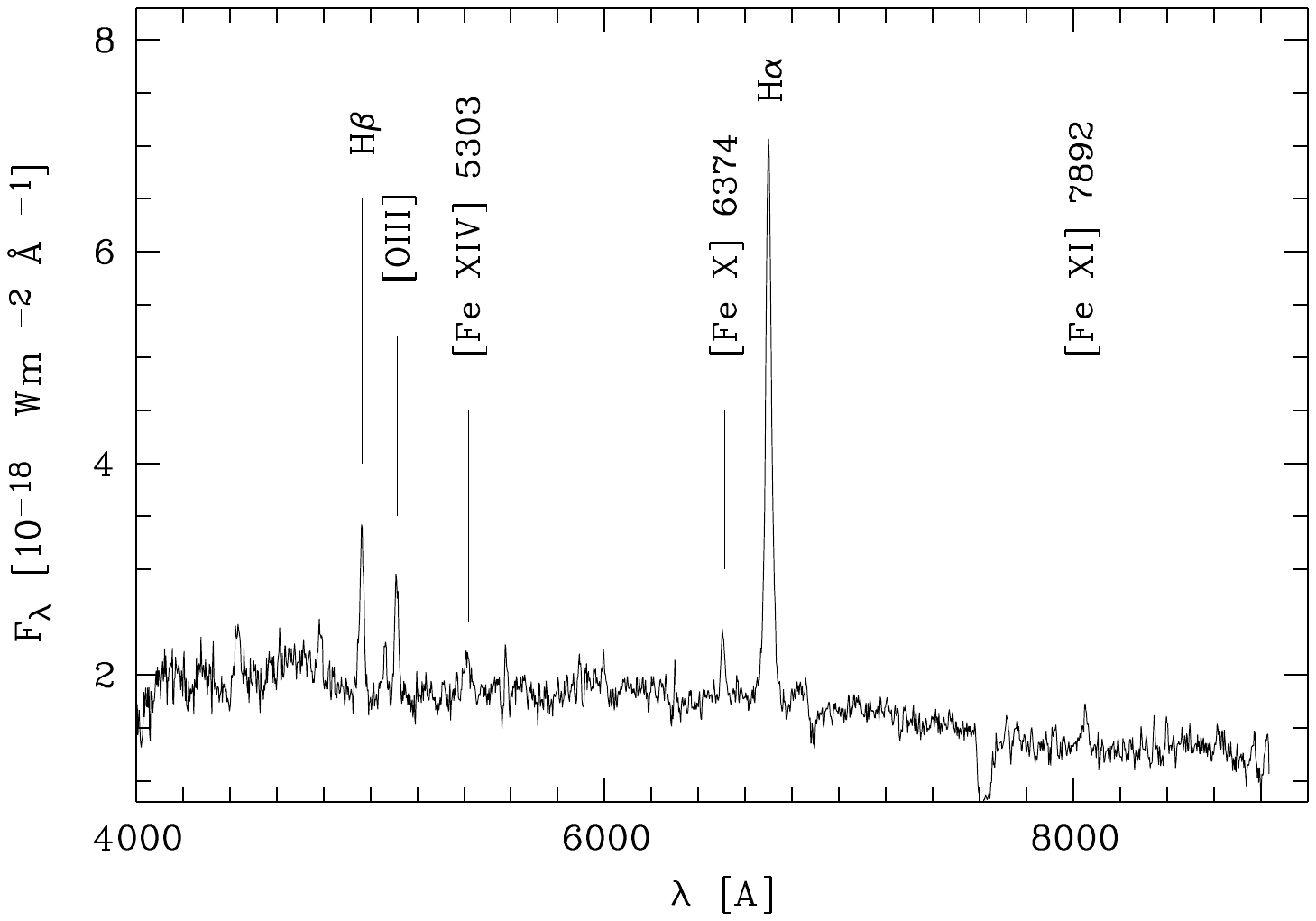}
    \caption{
    Xinglong optical spectrum of IC 3599. The brightest emission lines are marked. 
    }
\label{Xinglong}
\end{figure}

\end{document}